\documentclass[prl,aps,tighten,twocolumn]{revtex4}           
\usepackage{graphics}
\usepackage{graphicx}

\begin{document} 
%\draft

%\wideabs{

\title{
Are Halos of Collisionless Cold Dark Matter Collisionless?}

\author{Chung-Pei Ma and Michael Boylan-Kolchin}

\affiliation{Departments of Astronomy and Physics, University of
California at Berkeley, Berkeley, CA~94720}

\begin{abstract}   

Much recent discussion about dark matter has been centered on two
seemingly independent problems: the abundance of substructure in dark
matter halos, and the cuspiness of the halos' inner density profile.
We explore possible connections between the two problems by studying
the gravitational scattering effects due to subhalos on the
phase-space distribution of dark matter particles in the main halos.
Our series of controlled numerical experiments indicates that the
number and mass density of subhalos can be high enough to cause the
collisionless dark matter particles in the inner part of a main halo
to diffuse, flattening the main halo's inner cusp within a few
dynamical times.  Depending on the masses and concentration of the
subhalos, the inner density profile of the whole system (main plus sub
halos) can either steepen or flatten.  Subhalo accretion can therefore
introduce significant scatter in the inner density profiles of 
dark matter halos, offering a possible explanation for the range of
profiles seen in both observations and cosmological simulations.
\end{abstract}

\maketitle

Particles undergo random walks and diffusion through collisional
scatterings.  The most noted example is the Brownian motion of small
macroscopic particles, whose velocities exhibit frequent sudden
changes due to impulsive collisions with individual molecules in a
liquid.  On astrophysical scales, stars also undergo random walks in
velocity space due to gravitational scatterings with, e.g., other
stars or giant interstellar clouds in a galaxy \cite{star}.
%in a globular cluster 

Recent high resolution $N$-body simulations of hierarchical structure
formation in cold dark matter (CDM) models have shown that spatial
distribution of dark matter in galaxy-hosting halos is not entirely
smooth.  Instead, roughly 10\% of a halo's mass is in the form of
hundreds to thousands of smaller, dense satellite subhalos of varying
mass \cite{Moore01}.  In this Letter we examine whether these subhalos
can be the source of a fluctuating gravitational potential that
produces collisional transport of CDM particles in the main halo, even
when the self-interaction of CDM is {\it collisionless}.  Our approach is
based on numerical simulations and addresses the fully nonlinear
regime of halo-subhalo interaction; a complementary approach is
pursued by \cite{MB04}, who have used second-order cosmological
perturbation theory to derive a kinetic equation for the phase-space
distribution of halo dark matter particles.

%In this Letter we explore the possibility that individual
%collisionless dark matter particles can experience noticeable
%collisional relaxation through gravitational interactions with dark
%matter subhalos within a parent halo.  Recent high resolution $N$-body
%simulations of hierarchical structure formation in cold dark matter
%(CDM) models have shown that CDM in galaxy-hosting halos is not
%entirely smoothly distributed in space.  Instead, roughly 10\% of a
%halo's mass is in the form of hundreds to thousands of smaller, dense
%satellite subhalos of varying mass \cite{Moore01}.  We examine whether
%these subhalos provide the source of lumpiness for a fluctuating
%potential that produces collisional transport of CDM particles in the
%main halo, even when the self-interaction of CDM is collisionless.

%Our approach here is based on numerical simulations and addresses the
%fully nonlinear regime of halo-subhalo interaction.  A complementary
%analytic study \cite{MB04} discusses a kinetic equation for the
%phase-space distribution of halo dark matter particles derived from
%the second-order cosmological perturbation theory.

%\section{Physics of Diffusion}

{\it Physics of Diffusion.}---A test particle of mass $M_t$ and
velocity $\vec{v}_t$ experiences dynamical friction and exhibits
random walks (in velocity space) as it moves through the gravitational
potential of background particles of mass $M_b$.  Both processes
change the test particle velocity ($\Delta v_i, i=1,2,3$) and energy
($\Delta E$): the dynamical friction is described by the diffusion
coefficient $D(\Delta v_\parallel)$ where $d \vec{v}_t/dt=\hat{v}_t
D(\Delta v_\parallel)$; the random walk is described by the diffusion
tensor $D(\Delta v_i \Delta v_j)$.  The rate of change of the kinetic
energy of the test particle is $D(\Delta E)=M_t \Sigma_i [v_i\,
D(\Delta v_i) + {1\over 2}D(\Delta v_i^2)]$ \cite{BT}.  For background
particles with a uniform mass density $\rho_b$ and an isotropic
Maxwellian velocity distribution with dispersion $\sigma_b$, it is
\begin{equation}
   D(\Delta E) = 4\pi G^2 \frac{\rho_b M_t}{v_t} \ln\Lambda \left\{
      -M_t F(x) + M_b [{\rm erf}(x)-F(x)]\right\} \,,
\label{deltaE}
\end{equation}
where $\ln\Lambda$ is the Coulomb logarithm, $x=v_t/\sqrt{2}\sigma_b$,
and $F(x)={\rm erf}(x)-2x\,\exp(-x^2)/\sqrt{\pi}$.  Note that this
equation is valid for arbitrary ratios of $M_t/M_b$ \cite{BM92}.
The first term in Eq.~(\ref{deltaE}) describes the energy {\it loss}
of the test particle due to dynamical friction.  In the standard
Chandrasekhar picture, a large test mass $M_t$
scatters off a sea of small background particles with mass $M_b$. 
In this limit ($M_t \gg M_b$), the first term in Eq.~(\ref{deltaE})
(due to the test particle polarizing the background medium) dominates,
and the second term is typically ignored.

Our focus in this Letter is different.  We are interested in the
effects on the dark matter particles in the main halo (our test
particles) due to the ensemble of subhalos (our background particles).
The relevant mass range, $M_t \ll M_b$, is therefore opposite of that
in the last paragraph.  The polarization cloud term is completely
negligible.  Instead, the key process is the second term in
Eq.~(\ref{deltaE}), which describes the {\it heating} of the test
particle due to stochastic fluctuations in the background particles.
Changes in the potential due to the distribution of dark matter
substructure are the dominant scattering source in our study.

%\section{Effects of Diffusion: Numerical Simulations}

{\it Effects of Diffusion.}---We perform a series of fully dynamical
simulations using {\tt GADGET}, a publicly available N-body tree code
\citep{gadget}, to follow the evolution of dark matter in
a parent halo containing an ensemble of subhalos.  To study the
dynamical interplay between a main halo and its subhalos in a
controlled and semi-realistic way, we use subhalo properties
similar to those from earlier full-scale cosmological simulations
\cite{Moore01}.  This strategy allows us to perform a suite of
numerical experiments to quantify the effects due to a wider range of
subhalo masses, concentration, and orbits than is possible with large
cosmological simulations.

Initially the particles in the main halo are given an NFW radial
density profile \cite{NFW}:
$\rho(r)=\rho_{crit}{\bar\delta}/[x(1+x)^2]$, where $x=r/r_s,
{\bar\delta}=200c^3/3[\ln(1+c)-c/(1+c)]$, and the concentration
parameter $c=r_{vir}/r_s$ is the ratio of the halo's virial to scale
radius.  We use a total of $10^6$ simulation particles for the main
halo and a force softening of $0.015 r_s$.  The particle velocities
are drawn from a local isotropic Maxwellian distribution where the
radius-dependent velocity dispersion is computed from the Jeans
equation (see \cite{BKM04} for details and tests of a similar set-up).
This velocity setup may cause an NFW halo's inner cusp to artificially
flatten initially \cite{kaz}.  To work around this problem, we evolve
an NFW halo in isolation for $\sim 8 \, t_{dyn}$ (where $t_{dyn}^2
\equiv 3 \pi / 16 G \rho(r_s)$) and use this
evolved halo as our initial main halo.  Our tests have shown that this
halo, which does differ slightly at $r \le 0.1 r_s$ from its original
NFW structure, is extremely stable over the next $10 t_{dyn}$ at all
scales $r \ge 0.03 r_s$.
%We have performed test runs
%to ensure that the main halos set up in this equilibrium state suffer
%negligible numerical two-body relaxation down to $r\approx 0.05 r_s$
%(see \cite{BKM04} for details and tests of a similar set-up).

To simulate the effects of substructures on dark matter halos, we add
$\sim 1000$ subhalos to the main halo.  The subhalo masses are drawn
from $dn_{sub}/dM_{sub}\propto M_{sub}^{-1.7}$, similar to those found
in cosmological simulations \cite{Moore01}.  Initially the subhalos
are placed within the virial radius of the main halo either with a
top-hat or $r^{-2}$ radial number density distribution (see Fig.~1 for
a comparison).  The center-of-mass velocities of the subhalos are
drawn from a local isotropic distribution identical to that of the CDM
particles in the main halo.  Simulations indicate that both CDM
particles and subhalos are likely to develop mild velocity
anisotropies in the outer part of the main halo \cite{diemand} but
this effect should be small.

%\subsection{Point-Mass Subhalos}

{\it Point-Mass Subhalos.}---In the simplest case, we represent each
subhalo as a point mass.  This model is unrealistic and overestimates
the relaxation effect since it ignores mass losses due to tidal
stripping.  However, it serves as a test case for the validity of the
standard Chandrasekhar formula (see Eq.~[2]) and approximates the
effects of dense baryonic clumps that can survive into halo centers.

%%%%%%%%%%%%%%%%%%%%%%%%%
\begin{figure}
%\hbox to \hsize{\hss \epsfxsize=8.6cm \epsffile{rhom.ps}\hss} 
\includegraphics[scale=0.5]{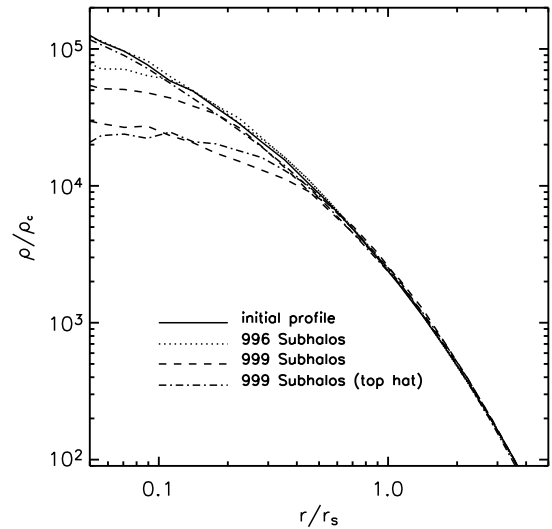}
\caption
{Evolution of the radial density profile of a main halo (with
$c_{main}=5.2$) containing 996 vs. 999 point-mass subhalos.  The
dotted (996) and dashed (999) curves are for identical simulations
except for the removal of the three most massive subhalos.  The inner
$\rho(r)$ decreases with time from $t=0$ (solid) to $2.8\,
t_{dyn}(r_s)$ (upper of each pair) and $6.9\,t_{dyn}(r_s)$ 
(lower of each pair);
%where $t_{dyn}$ is the dynamical time of themain halo at $r_s$ and 
$t_{dyn}(r_s)$ is 0.072, 0.25, and 0.45 $h^{-1}$ Gyr for a $10^8
M_\odot$ (at $z=4$), $10^{12} M_\odot$ ($z=1$), and $10^{14} M_\odot$
($z=0$) main halo.  The dot-dashed curves compare the same 999 run as
the dashed curves except the subhalo centers are placed initially with a
tophat instead of $r^{-2}$ distribution; the main halo here flattens
later between $6.9\,t_{dyn}$ (upper dot-dashed) and $9.7\,t_{dyn}$
(lower).  Without the subhalos, we have tested that the solid curve
does not change over at least $10\,t_{dyn}$.  }
\label{fig1}
\end{figure}
%%%%%%%%%%%%%%%%%%%%%%%%%

Fig.~1 shows the results of our point-mass subhalo simulations in
dimensionless units, which hold for halos of different masses at
appropriately scaled cosmic times (see caption).  It illustrates
that point-mass subhalos can indeed result in significant flattening
in the inner $\rho$ of a main halo within a few (inner) dynamical
times.
%($t_{dyn}^2 \equiv 3 \pi / 16 G \rho(r_s)$ unless otherwise noted).  
The amount of flattening is sensitive to the masses
of the several most massive subhalos present in the main halo since
these subhalos dominate the energy exchange with the dark matter in
the main halo, as seen in Eq.~(1).  We have performed two identical
runs -- one having 999 point subhalos (dashed curves); the other
having 996 point subhalos (dotted) without the top 3 most massive
subhalos in the 999 run -- to test the effect of massive subhalos.
The subhalos are placed initially within the main halo with $r^{-2}$
distribution.  Fig.~1 shows the heating of the main halo in the
996 run occurs at a later time (between 2.8 and $6.9\, t_{dyn}$) and
also leads to less flattening than the 999 run.  The total subhalo
mass in the 999 and 996 runs is 7.02\% and 3.29\%$M_{main}$;
the three most massive subhalos in the 999 run have
masses 1.51\%, 1.25\%, and 0.97\%$M_{main}$.

The two 999 subhalos runs in Fig.~1 illustrate the dependence of the
relaxation timescales on the subhalo spatial distribution.  In
accordance with Eq.~(1), the inner part of the main halo flattens more
quickly for the $r^{-2}$ case than the top-hat case.  For the latter,
the initial main halo $\rho(r)$ is unchanged through $6.9\,t_{dyn}$
and then flattens quickly in three $t_{dyn}$.  The stabilized main
halo profile (at $9.7\,t_{dyn}$; bottom dot-dashed) is similar to the
other 999 case, so the difference due to different subhalo spatial
distributions is mainly in the timescales.

{\it Puffy Subhalos.}---To model the subhalos more realistically, we
perform a series of simulations in which the 10 most massive subhalos
are given NFW profiles, while 989 lower mass subhalos (all $< 0.1$\%
$M_{main}$) are represented by point particles since they would suffer
little mass loss.  The total mass in subhalos in these cases is equal
to either 7.02 or 10.3\%$M_{main}$, where the ten most massive
subhalos comprise 5.2\% or 9.0\%.  The 7.02\% model has the same subhalo
mass spectrum as in the point-mass runs above, while the 10.3\% model
uses a different realization in which the three most massive subhalos
are 4.66, 2.09, and 1.00\%$M_{main}$.

%Fig.~2 shows $\rho(r)$ from our simulations with NFW subhalos.  Since
%the subhalos can now shed mass in complicated ways, Fig.~2 contrasts
%$\rho(r)$ computed from only the main halo particles (left) vs all
%particles (right).  Three combinations of subhalo concentration
%$c_{sub}$ and subhalo mass fraction are compared.  
Fig.~2 shows $\rho(r)$ from our simulations of NFW subhalos with three
combinations of subhalo concentration $c_{sub}$ and subhalo mass
fraction.  Since the subhalos can now shed mass in complicated ways,
we compute $\rho(r)$ both from the main halo particles only
(left) and from all particles (right).
Flattening in the
main halo $\rho(r)$ is seen for all three cases.  The amount of
flattening is more severe when a higher mass fraction of the system is
in subhalos, and when the subhalos have a higher $c_{sub}$ because
more centrally concentrated subhalos suffer less tidal mass loss as
they sink towards the main halo center.

%%%%%%%%%%%%%%%%%%%%%%%%%
\begin{figure}
%\hbox to \hsize{\hss \epsfxsize=8.6cm \epsffile{rhom.ps}\hss} 
\includegraphics[scale=0.6]{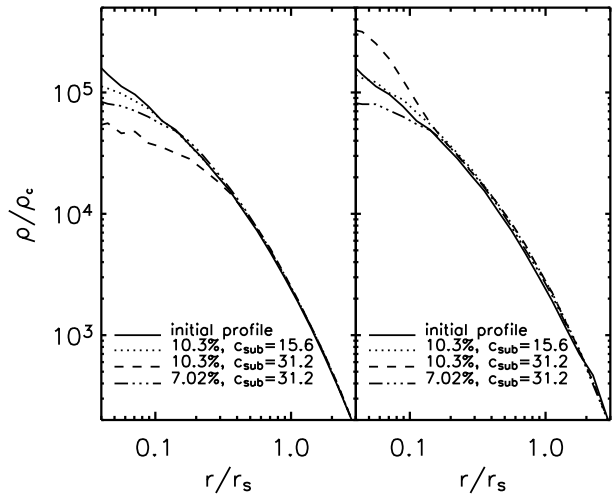}
\caption
{Radial density profile of a main halo (with $c_{main}=5.2$)
containing puffy subhalos.  The panels compare $\rho(r)$ of the main
halo (left) vs main-plus-subhalos (right) initially (solid) and after
$5.55\,t_{dyn}$ of evolution (other 3 curves).  Three simulations with
different subhalo concentration ($c_{sub}=15.6$ vs 31.2) and total
subhalo masses (10.3\% vs 7\% of main halo mass) are shown.} 
\label{fig2}
\end{figure}
%%%%%%%%%%%%%%%%%%%%%%%%%

%%%%%%%%%%%%%%%%%%%%%%%%%
\begin{figure}
%\hbox to \hsize{\hss \epsfxsize=8.6cm \epsffile{rhom.ps}\hss} 
\includegraphics[scale=0.6]{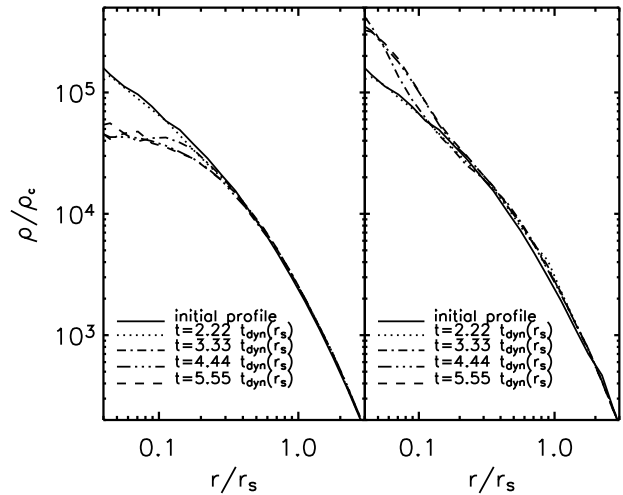}
\caption
{\small Time evolution of $\rho$ of the main halo (left) and
main-plus-subhalo (right) for the $c_{sub}=31.2$, 10.3\% subhalo run
in Fig.~2.  Most of the evolution occurs between 2.22 and
$3.33\,t_{dyn}$.}
\label{fig3}
\end{figure}
%%%%%%%%%%%%%%%%%%%%%%%%%

The inner cusp of the {\it total} mass, however, can steepen, remain
the same, or flatten, depending on the competition between the
addition from subhalo masses deposited in the central regions and the
removal of main halo particles due to gravitational heating.  The
three models in Fig.~2 (right panel) showcase the three outcomes.  The
subhalos in the 7.02\% model are not massive enough to add much mass,
so both the main and total $\rho(r)$ are flattened ($\sim r^{-0.75}$)
in $\sim 6\,t_{dyn}$.  In contrast, in the model with 10.3\% subhalo
mass and $c_{sub}=31.2$, the mass added by the most massive subhalos
(the top two have 4.66\% and 2.09\% $M_{main}$) more than compensate
for the flattening in the main halo, leading to a steeper than
$r^{-1}$ inner cusp.  Fig.~3 illustrates this model in more detail
with five time outputs: most of the evolution occurs within one
dynamical time after $2.2\,t_{dyn}$ when tje most massive subhalos
make their way to the center.  \cite{dekel} showed that accreting one
massive concentrated subhalo of 10\%$M_{main}$ can also produce a cusp
in an initially cored halo.

We emphasize that the inner $\rho(r)$ of a halo depends sensitively on
the location of the halo center used to compute $\rho(r)$.  Although
the initial momenta of the subhalos are drawn from an isotropic
distribution, fluctuations typically introduce a small center-of-mass
(COM) motion for the entire system: a COM velocity $\sim 2$\% of the
main halo circular velocity was not uncommon, resulting in COM offsets
of $\sim r_s$ in $\sim 10\,t_{dyn}$.  Neglecting this effect and
na\"{\i}vely using a halo's initial center as the center for
subsequent outputs would lead to a flattened $\rho(r)$.  We use a more
physical halo center (e.g. iteratively determined COM, most bound
particle, or COM of the 500 most bound particles; all three give
nearly identical results), which eliminates this spurious flattening.

{\it Timescale.}---How do the timescales seen in the simulations
compare with the simple energy exchange time predicted by
Eq.~(\ref{deltaE})?  The latter predicts
%timescale for the CDM particles ($M_t$) to gain energy from background
%subhalos ($M_b$)
\begin{equation}
   t_{relax} = \frac{{1\over 2}M_t v^2_t}{\left| D(\Delta E) \right|} 
   \approx \frac{1}{8\pi G^2 \ln \Lambda} 
	    \frac{v_t^3}{\rho_b M_b} \frac{\sqrt{\pi}}{2x e^{-x^2}} \,,
\label{trelax}
\end{equation}
where the second equality assumes $M_t \ll M_b$.  In our study, this
gives the timescale for heating the dark matter particles from a
background of subhalos of mass $M_{sub}$, density $\rho_{sub}$ in the
main halo, and COM velocity dispersion $\sigma_{sub}$.  Re-expressing
it in terms of the main halo's virial mass $M_{main}$ and radius $r_v$, and
1-d velocity dispersion at $r_v$, $\sigma(r_v)$, we obtain
\begin{equation}
    t_{relax} \approx \frac{0.12}{H(z)} \frac{10}{\ln\Lambda} 
    \frac{\rho_{crit} M_{main}}{ \rho_{sub} M_{sub}} 
	    \left( \frac{v_t}{\sigma(r_v)} \right)^3
          \frac{\sqrt{\pi}}{2x e^{-x^2}} \,,
\end{equation}
where the Hubble time at redshift $z$ is $H^{-1}(z)=9.78\, {\rm Gyr}\,
h^{-1}[\Omega_m(1+z)^3 +\Omega_\Lambda]^{-1/2}$.  Let
$dn_{sub}/dM_{sub}\propto M_{sub}^{-\alpha}$ be the subhalo mass
function (assuming $\alpha > 1$), $\beta$ be the ratio of the total
mass in subhalos to $M_{main}$, and $\gamma$ be the ratio of the most
massive subhalo to $M_{main}$.  We find $\rho_{crit}
M_{main}/\rho_{sub} M_{sub}=(3-\alpha)/(2-\alpha)/(200\beta\gamma)$;
for the 999 point-mass model shown in Fig.~1, this ratio is about 19.
The local dynamical time at the scale and virial radius of an NFW halo
with $c_{main}=5.2$ is
$t_{dyn}(r_s)=0.14\,t_{dyn}(r_v)=0.046\,H^{-1}(z)$ (assuming unity for
factors involving velocities, which is likely an underestimate), so we
find $t_{relax}\sim 2.3\,H^{-1}(z) \sim 50\,t_{dyn}(r_s)\sim
7\,t_{dyn}(r_v)$.  This is at least 5 times longer than the flattening
timescale seen in the 999 point-mass top-hat simulation in Fig.~1.
Eqs.~(1)-(3), however, are valid only for a stationary, infinite,
homogeneous background with a global Maxwellian velocity distribution
\cite{BM92}.  In our study, the background is an ensemble of dark
matter subhalos, themselves moving in a deeper main halo potential and
experiencing dynamical friction and tidal mass losses.  While
Eqs.~(1)-(3) elucidate the energy exchange between subhalos and dark
matter particles, it is not surprising that they do not predict the
exact timescales seen in simulations.

We have performed a test run with 1000 point-mass subhalos of equal
mass where the total subhalo mass is 15\% $M_{main}$.  This subhalo
mass spectrum is unrealistic, but this run provides an additional test
case and a comparison case for a recent cluster galaxy study
\cite{elzant}.  Since each subhalo mass in our test run is only
0.015\% $M_{\rm main}$, $\sim 100$ times smaller than the most massive
subhalos in Fig.~1, $t_{relax}$ in Eq.~(3) increases by a factor of
$\sim 100$, too long to result in change in the inner halo profile.
We indeed did not see any flattening over $\sim 9\,t_{dyn}$ (at $r_s$)
in our simulation.

%\section{Implications}

{\it Implications.}---Our series of controlled numerical experiments
indicates that collisionless dark matter particles in the inner parts
of galaxy and cluster halos can gain energy through gravitational
scatterings off concentrated dark matter subhalos, altering the inner
density cusp of the main halo within a few dynamical times.  These
subhalos appear ubiquitous in high resolution cosmological simulations
and provide the source of fluctuations for the diffusion described by
Eq.~(\ref{deltaE}).
We have studied the evolution of halos under the influence of only one
generation of subhalos, while real halos grow continuously by
accretion and mergers.  The effects we have seen, however, suggest
that fluctuations due to subhalos in parent halos are important for
understanding the time evolution of dark matter density profiles and
the halo-to-halo scatter of the inner cusp seen in recent ultra-high
resolution cosmological simulations \cite{navarro}.  We have shown
that this scatter may be explained by subhalo accretion histories:
when we allow for a population of subhalos of varying concentration
and mass, the total inner profile of dark matter can either steepen or
flatten.  

Recent observations of dwarf galaxy rotation curves based on
CO and H$\alpha$ emission find significant variations in the inner
profile, ranging from a core to $\sim r^{-1}$ \cite{blitz}.  While
baryonic physics can influence the central mass profile, the purely
gravitational physics of subhalo scattering studied here may also
accommodate the variations seen in these observations.
Conversely, maintaining a stable and universal inner profile would
require a ``quiescent'' accretion history not involving concentrated
massive subhalos.  Halos in cosmological models with truncated
small-scale power (e.g. warm dark matter) contain many fewer
satellites \cite{bode}; their inner cusps should therefore be less
prone to variations due to subhalo accretion.

We thank J. Arons, E. Bertschinger, A. Dekel, N. Gnedin and A.
Kravtsov for useful discussions.  The simulations were
performed at NERSC.
%the National Energy Research Scientific Computing Center.
%This research used resources of the National Energy
%Research Scientific Computing Center, which is supported by the Office
%of Science of the U.S. Department of Energy under Contract
%No. DE-AC03-76SF00098.  
C.-P. M is partially supported by a Cottrell Scholars Award from the
Research Corporation and NASA grant NAG5-12173.

\end{document}